\documentclass[twoside,twocolumn,english,pre]{revtex4}
\usepackage[T1]{fontenc}
\usepackage[latin1]{inputenc}
\usepackage{amsmath}
\usepackage{graphicx}
\usepackage{amssymb}

\makeatletter
\usepackage{babel}
\makeatother
\begin{document}

\preprint{ePrint four-page version III}

\title{Wavevector resonance in a nonlinear multi-wavespeed chaotic billiard}

\author{Alexei Akolzin}

\author{Richard L. Weaver}

\affiliation{Department of Theoretical and Applied Mechanics, University of Illinois,
104 S. Wright Street, Urbana, Illinois 61801, USA}

\begin{abstract}
Nonlinear coupling between eigenmodes of a system leads to spectral
energy redistribution. For multi-wavespeed chaotic billiards the average
coupling strength can exhibit sharp discontinuities as a function
of frequency related to wavevector coincidences between constituent
waves of different wavespeeds. The phenomenon is investigated numerically
for an ensemble of 2D square two-wavespeed billiards with rough boundaries
and quadratic nonlinearity representative of elastodynamic waves.
Results of direct numerical simulations are compared with theoretical
predictions.
\end{abstract}

\pacs{05.45.Mt, 43.25.+y, 43.35.+d, 62.30.+d}

\date{\today}

\maketitle
Recent years have been marked by increased attention to multiple scattering
and propagation of classical waves in disordered nonlinear media.
Main research has been in optics; see \cite{ref:optics.theoretical}
and references therein. Characteristics of a wave-field pertinent
to continuous-wave problems, such as angular correlations and coherent
backscattering, have dominated \cite{ref:optics.cbs}. Influence of
nonlinearity on the stability of a speckle pattern has been considered
also \cite{ref:optics.stability}; nonlinear phenomena in transient
fields, as, for example, frequency shifting of a lasing-mode in a
random laser \cite{ref:optics.temporal}, have been studied. 

An anisotropic or multi-wavespeed nature of the medium supporting
propagation of the classical waves allows a broader ground for interplay
between nonlinearity and multiple scattering. Statistical effects
of nonlinear wave-field behavior have received little attention so
far. A class of systems in which these effects are expected, is given
by elastodynamic waves in solids, with anisotropy or multiple speeds
of propagation as a rule rather than exception. Ballistic billiards
in the form of elastic solids with a ray-chaotic shape are conveniently
realizable and allow ready access to the time domain and observation
of transients \cite{ref:elastodynamics}. These billiards are representative
of classical wave-bearing systems, and are suited for the study of
statistical nonlinear effects that stem from the multiple-wavespeed
character of the wave-field. The purpose of this study is to provide
evidence of one of such effects, namely wavevector resonance in a
chaotic nonlinear billiard, with an elastodynamic billiard chosen
as an underlying physical model. 

Dynamics of an elastodynamic billiard after excitation by a transient
source can be reduced in the absence of energy dissipation to a set
of nonlinearly coupled oscillators with natural frequencies $\omega_{k}$,
amplitudes of vibration $d_{k}$, and coupling matrices $N$ \cite{ref:nonlinear}:\begin{equation}
\ddot{d_{k}}+\omega_{k}^{2}d_{k}+N_{klm}d_{l}d_{m}+N_{klmn}d_{l}d_{m}d_{n}=0,\label{eq:ODE}\end{equation}
 where nonlinearity up to cubic terms has been accounted for. Associated
with each oscillator (mode) is its linear energy $E_{k}=\left(\dot{d}_{k}^{2}+\omega_{k}^{2}d_{k}^{2}\right)/2$.
Although not being true energy in the presence of nonlinearity, $E_{k}$
elucidates trends of modal energy redistribution when nonlinearity
is weak.

For observation times much smaller than the inverse modal spacing,
$t\ll D$, individual modes are not resolved, and statistical description
is in order. One of the energy quantities conveniently accessible
for experimental measurement under these conditions is the average
spectral density, $E\left(\omega_{k},t\right)=D\left(\omega_{k}\right)\left\langle E_{k}\left(t\right)\right\rangle $,
a constant in the absence of dissipation or nonlinearity. Under the
influence of nonlinearity it evolves in time. The evolution leads
to energy deposit into frequencies that contained no energy initially,
thereby allowing detection of the nonlinearity. In case of a weak
nonlinearity the leading contribution to the deposit is given by convolution
of energy densities at two frequencies that have a given (target)
frequency $\omega$ as a combination, \emph{i.e.} their sum or difference
\cite{ref:nonlinear}: \begin{align}
\dot{E}\left(\omega,t\right)=D\left(\omega\right) & \int_{0}^{+\infty}d\omega'\sum_{\pm}\mathbb{N}\left(\omega,\omega',\left|\omega\pm\omega'\right|\right)\nonumber \\
\times & E\left(\left|\omega\pm\omega'\right|\right)E\left(\omega'\right)/\left(\omega\pm\omega'\right)^{2}{\omega'}^{2}.\label{eq:power}\end{align}
 Redistribution of the energy involving triads of frequencies is characteristic
of a dominant quadratic nonlinearity \--- contribution of the cubic
terms of eq. (\ref{eq:ODE}) vanishes on the average, being of the
next order of smallness, \--- and is in agreement with behavior expected
from elementary theory of nonlinear oscillations (cf., for example,
\cite{ref:Mierovitch}). It corresponds to internal (frequency) resonance
between the modes of the billiard, which effectively takes place when
individual modes are not resolved, and hence a combination frequency
of two source modes is indistinguishable from the frequency of another
mode of the system. The resonance manifests itself in early-time linear
energy growth that can be linked to secular terms arising in Regular
Perturbation Theory \cite{ref:nonlinear}. 

Redistribution of the spectral energy in the billiard occurs due to
nonlinear coupling of the modes. The average coupling strength is
provided by the function, \begin{align}
\mathbb{N} & \left(\omega_{k},\omega_{l},\omega_{m}\right)=\left(\pi/2\right)\left\langle N_{\underline{k}\underline{l}\underline{m}}N_{\underline{k}\underline{l}\underline{m}}\right\rangle \nonumber \\
 & =\left(\pi/2\right)\hat{N}_{\alpha\beta\gamma}\hat{N}_{\nu\mu\eta}\; K_{\alpha\nu}\left(\omega_{k}\right)K_{\beta\mu}\left(\omega_{l}\right)K_{\gamma\eta}\left(\omega_{m}\right).\label{eq:N}\end{align}
 Greek indices denote spatial degrees of freedom including both spatial
position $\mathbf{x}$ and Cartesian indices, and repeated indices
imply summation. The function is symmetrical with respect to its arguments,
rendering interaction strength of any triad of frequencies independent
of energy transfer direction between them. Nevertheless, in nondispersive
billiards with the inverse modal spacing given by Weyl series, $D\sim V_{d}\omega^{d-1}+\ldots$
\cite{ref:Gutzwiller}, where $V_{d}$ stands for $d$-dimensional
volume of the billiard, the overall power input exhibits a trend of
the energy to be transfered up the frequency spectrum.

Specifics of the physics of a particular system reside in statistics
of its modes $u_{k}$, and operator $\hat{N}$ that acts on them,
$N_{klm}=\hat{N}_{\alpha\beta\gamma}u_{k}\left(\alpha\right)u_{l}\left(\beta\right)u_{m}\left(\gamma\right)$.
Under the assumption that the modes are mutually uncorrelated Gaussian
random vectors \cite{ref:RWM,ref:Gaussian}, the statistics are fully
provided by the pairwise spatial modal correlator $K_{\alpha\beta}\left(\omega_{k}\right)=\left\langle u_{\underline{k}}\left(\alpha\right)u_{\underline{k}}\left(\beta\right)\right\rangle $.
In the short-wavelength limit, when boundary effects are unimportant,
the correlator can be approximated by means of the Green's function
in boundless medium, with long-range correlation being limited to
billiard diameter $L$. Refined approximations of the correlator that
include finite size and geometry effects can be constructed upon need,
see \cite{ref:Urbina} and references therein.

Under a Random Wave Model (RWM) a mode inside a chaotic billiard can
be viewed as superposition of constituent plane waves coming from
all directions and having random amplitudes and phases \cite{ref:RWM}.
In a multi-wavespeed billiard these waves can have different speeds
of propagation and different amplitudes, requiring a modification
to RWM \cite{ref:GBerry}. Each pair of the constituent waves from
the source modes interacts nonlinearly producing a constituent wave
of the target mode. Two conservation laws must be obeyed in the process.
The first requires frequency of the resultant wave to be equal either
to the sum or difference of the source ones. A parallel can be drawn
to energy conservation applied to particle scattering problems. The
law is automatically satisfied by the internal-resonance structure
of eq. (\ref{eq:power}). The second law requires wavevectors of the
constituent waves to sum, thus standing for conservation of wave pseudo-momentum.
In nondispersive systems interaction of constituent waves of the same
wavespeed is always allowed by the above conservation laws, when the
waves are collinear. This mechanism provides a non-zero background
coupling strength for any frequency combination of source modes. Interaction
of constituent waves of different wavespeeds, however, is only possible
if the frequencies (and polarizations) of the source waves are in
special relation to each other \cite{ref:phonons}. If the frequencies
are such, this additional interaction channel is open, and nonlinear
coupling strength is expected to be greater. Integral contribution
of all constituent-wave interactions is provided by eq. (\ref{eq:N}),
statistics of the waves being incorporated by means of spatial correlator
$K$. Precise conditions under which the maximum of the coupling strength
is achieved depend on the specifics of the nonlinearity; for typical
elastic solids they were seen to imply interaction of the constituent
plane waves in a nearly collinear fashion \cite{ref:nonlinear}. By
analogy to internal (frequency) resonance a sharp increase in the
coupling strength due to nonlinear interaction of constituent waves
with different wavespeeds is termed wavevector resonance here.

To verify existence of the wavevector resonance a series of direct
numerical simulations (DNS) was performed. An ensemble of discrete
2D square billiards with average boundary roughness of one twenty-fourth
of the billiard size was taken; cf. \cite{ref:Weaver.DNS}. Evolution
of the wave-field inside the billiards was governed by first-order
finite-difference version of the following two-wavespeed elastodynamic
equation:\begin{equation}
\ddot{u}_{i}=c_{t}^{2}u_{i,jj}+\left(c_{l}^{2}-c_{t}^{2}\right)u_{j,ij}+\varepsilon\Phi_{ijklmn}\left(u_{k,l}u_{m,n}\right)_{,j},\label{eq:dns}\end{equation}
 where $c_{l}$ and $c_{t}$ are longitudinal and transverse wavespeeds,
respectively. The nonlinear coupling term was chosen to be quadratic
in derivatives of the displacements $u$, a form representative of
the physical nonlinearity in isotropic elastic solids described by
the five-constant theory \cite{ref:Ogden}. The derivatives were coupled
by the elementary isotropic tensor, \begin{align*}
\Phi_{ijklmn} & =\\
\frac{1}{8}\bigl( & \delta_{ik}\delta_{jm}\delta_{ln}+\delta_{ik}\delta_{jn}\delta_{lm}+\delta_{il}\delta_{jm}\delta_{kn}+\delta_{il}\delta_{jn}\delta_{km}\\
+ & \delta_{im}\delta_{jk}\delta_{ln}+\delta_{im}\delta_{jl}\delta_{kn}+\delta_{in}\delta_{jk}\delta_{lm}+\delta_{in}\delta_{jl}\delta_{km}\bigr).\end{align*}
 Strength of the nonlinearity was controlled by parameter $\varepsilon$,
which was kept small on the order $10^{-2}$, while $u=O\left(1\right)$.
Dirichlet boundary conditions were imposed on the billiard boundaries.
Initial displacement field was arranged so that only modes supporting
two narrow-band Gaussian peaks of the spectral energy centered at
given frequencies $\omega_{1}$ and $\omega_{2}$ were excited. The
width of the peaks was taken wide enough to encompass tens of modes.
Evolution of the initial field up to one tenth of the Heisenberg time,
$t_{H}=\left(A/2\right)\left(c_{l}^{-2}+c_{t}^{-2}\right)c_{t}/h$,
where $A$ is average billiard area, and $h$ is finite-difference
step length, was then computed by directly solving the governing ODEs
(\ref{eq:dns}). The procedure was performed for a number of realizations
in order to obtain ensemble average of the spectral energy. Expected
linear growth in time of the energy peaks centered at combinatoric
frequencies, $\omega_{1\pm2}=\left|\omega_{1}\pm\omega_{2}\right|$,
was observed. By varying central frequencies of the source peaks,
and calculating average energy growth rate of the combinatoric peaks,
the coupling strength (\ref{eq:N}) was recovered \cite{ref:nonlinear}:\begin{equation}
\mathbb{N}\left(\omega_{1\pm2},\omega_{1},\omega_{2}\right)=\left(1/2\right)\dot{E}_{1\pm2}\omega_{1}^{2}\omega_{2}^{2}/D\left(\omega_{1\pm2}\right)E_{1}E_{2},\label{eq:N.dns}\end{equation}
 here $E_{1,2}$ and $E_{1\pm2}$ are total energies carried by the
source and combinatoric peaks, respectively. Theoretical estimate
of $\mathbb{N}$ was also obtained by means of eq. (\ref{eq:N}) for
the purpose of comparison with the DNS. Nonlinear operator containing
specifics of nonlinearity model was taken in accordance with the one
used in eq. (\ref{eq:dns}):\begin{align*}
\hat{N} & {}{}_{\alpha=\left\{ \mathbf{x},i\right\} \beta=\left\{ \mathbf{x}',k\right\} \gamma=\left\{ \mathbf{x}'',m\right\} }\\
 & =\varepsilon\Phi_{ijklmn}\delta\left(\mathbf{x}-\mathbf{x}'\right)\delta\left(\mathbf{x}-\mathbf{x}''\right)\partial^{3}/\partial x_{j}\partial{x'}_{l}\partial{x''}_{n}.\end{align*}
 Short-range behavior of the spatial modal correlator was approximated
by its behavior in boundless medium, \begin{align}
K^{\infty} & {}_{\alpha=\left\{ \mathbf{x},i\right\} \beta=\left\{ \mathbf{x}+\mathbf{r},j\right\} }=A^{-1}\left[c_{l}^{-2}+c_{t}^{-2}\right]^{-1}\nonumber \\
\times & \biggl\{\left(\delta_{ij}/2\right)\left[c_{l}^{-2}J_{0}\left(k_{l}r\right)+c_{t}^{-2}J_{0}\left(k_{t}r\right)\right]\nonumber \\
 & +\left(\delta_{ij}/2-\hat{r}_{i}\hat{r}_{j}\right)\left[c_{l}^{-2}J_{2}\left(k_{l}r\right)-c_{t}^{-2}J_{2}\left(k_{t}r\right)\right]\biggr\}.\label{eq:K}\end{align}
 Area of the billiard $A$ enters the above expression due to normalization
of the modes to unity. The short-range correlator, as it is, leads
to divergence of the integrals at large separations in eq. (\ref{eq:N}).
To account for finite domain size and to keep integrals convergent
spatial correlation of the modes was restricted by billiard diameter
$L$: $K=K^{\infty}\exp\left(-r/L\right)$. The coupling strength
was then found to scale asymptotically as $\sqrt{L}$ with linear
system size. The scaling is an artifact of problem dimensionality,
and is not found in 3D, where $\mathbb{N}$ remains $O\left(L^{0}\right)$
\cite{ref:nonlinear}. With non-trivial dependence of the coupling
strength on the system size, specifics of the average domain shape
become important. They could systematically be accounted for, but
such undertaking lies beyond the scope of the present work \cite{foot:LRC}.

\begin{figure}
\begin{center}(a)\includegraphics{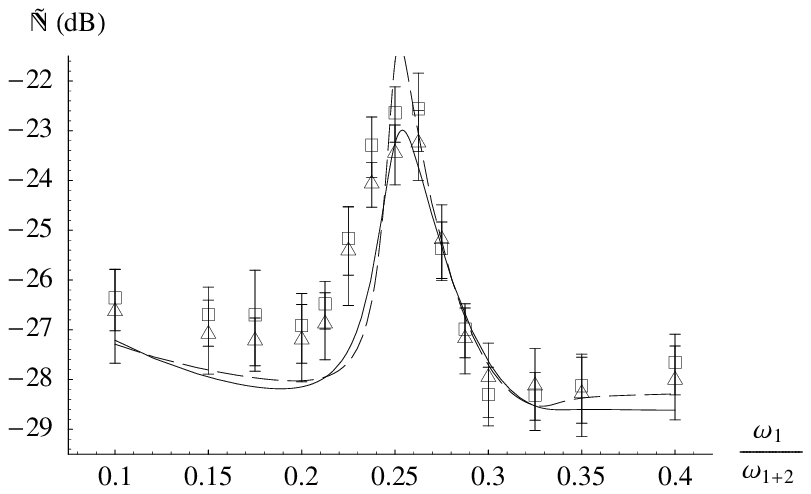}\end{center}

\begin{center}(b)\includegraphics{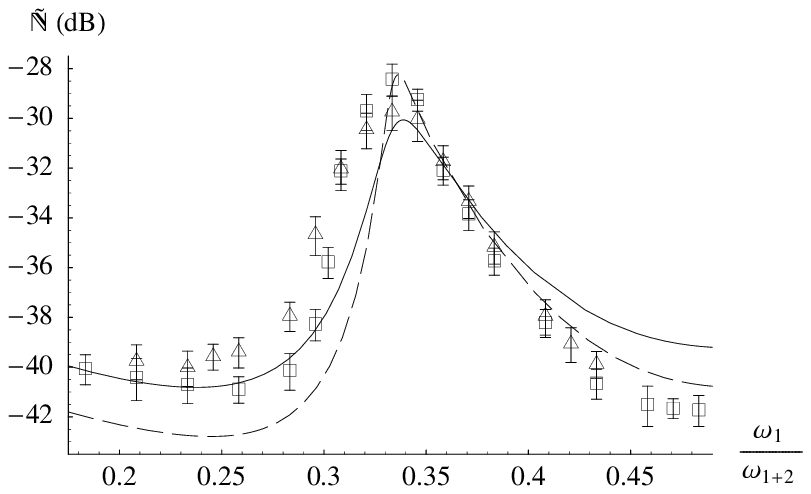}\end{center}

\caption{\label{cap:N} Normalized coupling strength for (a) $c_{l}/c_{t}=2$,
$\omega_{1+2}=1.2\, c_{t}/h$; and (b) $c_{l}/c_{t}=3$, $\omega_{1+2}=1.0\, c_{t}/h$.
Solid and dash lines represent theoretical estimates, symbols $\vartriangle$
and $\square$ provide DNS data for nominal system sizes of $128$
and $256$ grid points, averaged over $25$ and $10$ realizations,
respectively. Error bars correspond to one standard deviation.}
\end{figure}

Comparison of the DNS results with theoretical estimates is given
in Fig. \ref{cap:N}. %
 Energy growth rate at the sum frequency was utilized to obtain numerical
values of the coupling strength; calculations at the difference frequency
were carried out as well to check proposed symmetry of $\mathbb{N}$.
To factor out dependence of the coupling strength on global parameters,
and isolate behavior associated with frequency interrelation of the
coupled modes, its normalized version was deemed best suited for analysis:
$\tilde{\mathbb{N}}\left(\omega_{1}/\omega_{1+2},k_{1+2}L\right)=\pi^{-2}\varepsilon^{-2}A^{2}\left[1+\left(c_{l}/c_{t}\right)^{-2}\right]^{3}\left(k_{1}k_{2}k_{1+2}\right)^{-4/3}\mathbb{N}$,
with $k=\omega/c_{t}$. Plotted in this form the coupling strength
reveals its increase at the resonance frequency ratios in the vicinity
of $\omega_{1}/\omega_{1+2}=\left(1\pm c_{t}/c_{l}\right)/2$. The
ratios correspond to collinear interaction of two constituent transverse
waves producing a longitudinal one. They define a lower and upper
bound of the source to target frequency ratio of the coupled modes
for which the wavevectors of the constituent waves can sum, \emph{i.e.}
their interaction is possible in accordance with pseudo-momentum conservation
law. Another constituent-wave interaction possible in physical solids,
with $c_{l}/c_{t}>1$, involves a longitudinal and a transverse wave
producing a longitudinal wave. Collinear wavevector summation for
this interaction type occurs at $\omega_{1}/\omega_{1+2}=\left[1\pm\left(3-c_{l}/c_{t}\right)/\left(1+c_{l}/c_{t}\right)\right]/2$.
However, this resonance does not manifest itself for the given nonlinearity
model. Under normal conditions it is expected to be noticeably smaller
than the one involving two transverse source waves due to equipartition
of the energy carried by two different types of constituent waves
inside a solid. Since most of the energy, in particular, a fraction
$\left(c_{l}/c_{t}\right)^{2}>1$, is carried by transverse waves,
their participation ratio in the coupling strength is expected to
be higher than that of longitudinal ones.

Although good qualitative, if not quantitative, agreement between
theoretical estimates and DNS was found for parameters used, several
remarks on their discrepancies ought to be made. A difference between
background off-resonance coupling strength levels, especially at low
frequency ratios, is noted. The difference can be attributed to the
fact that though theoretical predictions based on eq. (\ref{eq:N})
are fit-parameter free, some ambiguity in defining effective volume
(2D area) of the billiard interior, where nonlinear mode coupling
takes place, exists. In the present theoretical estimates it was taken
exactly equal to the average domain area. No account of boundary and
confinement effects was made by the assumed correlator form (\ref{eq:K})
with qualitative long-range correlation dependence on the order of
domain size. Near the boundaries increased correlation of the modes
due to boundary conditions is found, greater coherence in their interaction
and corresponding increase in the observed coupling strength is expected.
Such effects can play a noticeable role for system sizes within current
DNS reach. In addition, at low frequency ratios wavelength of the
lowest source mode becomes large with respect to boundary roughness,
and comparable to domain size. Regular structure of the mode acquired
in this case leads to deviation from the assumed modal statistics,
and is expected to provide increased coherence of mode coupling in
the interior of the billiard. Large-scale symmetry of the square billiard,
however, seems to be of less significance in this case, as DNS calculations
of the coupling strength for ensemble of rough-boundary 3:2 aspect-ratio
quarter-stadium billiards produced compatible results. 

\begin{figure}
\begin{center}\includegraphics{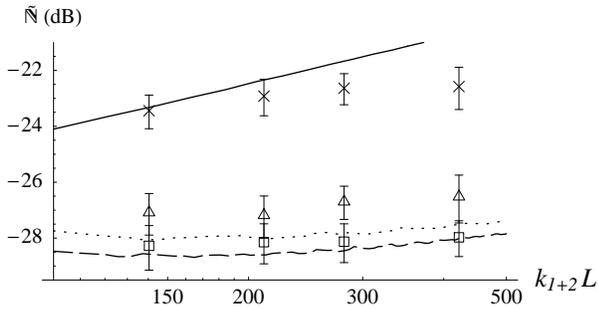}\end{center}

\caption{\label{cap:scaling} Normalized coupling strength for $c_{l}/c_{t}=2$,
$\omega_{1+2}=1.2\, c_{t}/h$. Dot, solid and dash lines represent
theoretical estimates at $\omega_{1}/\omega_{1+2}=0.15$, $0.25$
and $0.35$. Symbols $\vartriangle$, $\times$ and $\square$ provide
DNS data for nominal system sizes of $128$, $192$, $256$ and $384$
grid points, averaged over $25$, $25$, $10$ and $10$ realizations,
respectively.}
\end{figure}

As mentioned above, coupling strength is predicted to asymptotically
grow as $\sqrt{L}$ with system size. The prediction, however, is
not supported by the DNS data, see Fig. \ref{cap:scaling}. %
 The origins and implications of the disagreement are not fully understood
at the time. However, it is speculated that it can be the result of
frequency smoothing imposed on the coupling function by the finite
width of the source energy distribution. The smoothing leads to the
fact that only $\mathbb{N}$, averaged over characteristic peak width,
could be recovered with the help of eq. (\ref{eq:N.dns}). Both decrease
and broadening of the mode coupling resonance peak will follow from
this smoothing. System sizes currently available for DNS realizations
do not allow direct observation of the expected $\sqrt{L}$ scaling,
or draw a conclusive statement regarding its absence. Decrease of
the source energy distribution width would be problematic, in that
it needs to contain many modes in order for statistical approach to
be valid.

In summary, existence of the wavevector resonance in the mode coupling
strength, responsible for nonlinear redistribution of the average
spectral energy, was verified by direct numerical simulation in a
two-wavespeed chaotic billiard with nonlinearity representative of
isotropic elastic solids. Qualitative agreement with theoretical estimates
in the absence of fit parameters was observed. Location of the coupling
strength peak was found to correspond to collinear interaction of
constituent waves of different wavespeeds. The location was determined
from arguments based on conservation principles, and to leading order
depends solely on linear system quantities.

\begin{acknowledgments}
This work was supported in part by the National Science Foundation
grant CMS-0201346. 
\end{acknowledgments}


\begin{thebibliography}{10}
\bibitem{ref:optics.theoretical}S.E. Skipetrov and R. Maynard, in \emph{Wave scattering in complex
media}, edited by B.A. van Tiggelen and S.E. Skipetrov (Kulwer Academic
Publishers, 2003)
\bibitem{ref:optics.cbs}R. Bressoux and R. Maynard, in \emph{Waves and imaging through complex
media}, edited by P. Sebbah (Kulwer Academic Publishers, 2001); T.
Wellens, B. Gr\'{e}maud, D. Delande, and C. Miniatura, Phys. Rev.
E \textbf{71}, 055603(R) (2005)
\bibitem{ref:optics.stability}S. E. Skipetrov, J. Opt. Soc. Am. B 21, 168 (2004)
\bibitem{ref:optics.temporal}B. Liu, A. Yamilov, Y. Ling, J.Y. Xu, and H. Cao, Phys. Rev. Lett.
\textbf{91}, 063903 (2003)
\bibitem{ref:elastodynamics}R. L Weaver, in \emph{Waves and imaging through complex media}, edited
by P. Sebbah (Kulwer Academic Publishers, 2001)
\bibitem{ref:nonlinear}A. Akolzin and R. L. Weaver, Phys. Rev. E \textbf{69}, 066605 (2004)
\bibitem{ref:Mierovitch}L. Mierovitch, \emph{Elements of vibration analysis} (McGraw-Hill,
New York, 1986), Chap. 10
\bibitem{ref:Gutzwiller}M. C. Gutzwiller, \emph{Chaos in classical and quantum mechanics}
(Springer-Verlag, New York, 1990), Sec. 16.2
\bibitem{ref:RWM}M.V. Berry, J. Phys A \textbf{10}, 2083 (1977); P. O'Connor, J. Gehlen,
and E.J. Heller, Phys. Rev. Lett. \textbf{58}, 1296 (1987)
\bibitem{ref:Gaussian}S. W. McDonald and A. N. Kaufman, Phys. Rev. A \textbf{37}, 3067 (1988)
\bibitem{ref:Urbina}J. D. Urbina and K. Richter, J. Phys. A \textbf{36}, L495-L502 (2003);
Phys. Rev. E \textbf{70}, 015201(R) (2004)
\bibitem{ref:GBerry}K. Schaadt, T. Guhr, C. Ellegaard, and M. Oxborrow, Phys. Rev. E \textbf{68},
036205 (2003); A. Akolzin and R.L. Weaver, Phys. Rev. E \textbf{70},
046212, (2004) 
\bibitem{ref:phonons}G. L. Jones and D. R. Kobett, J. Acoust. Soc. Am. \textbf{35}, 5 (1963)
\bibitem{ref:Weaver.DNS}R.L. Weaver and J. Burkhardt, J. Acoust. Soc. Am. \textbf{95}, 3186
(1994); R.L. Weaver and O.I. Lobkis, J. Sound Vib. \textbf{231}, 1111
(2000)
\bibitem{ref:Ogden}R.W. Ogden, \emph{Nonlinear Elastic Deformations} (Ellis Horwood Ltd.,
Chichester, 1984), Sec. 6.1.6
\bibitem{foot:LRC}Terms associated with spatial derivatives of the long-range correlation
factor were also neglected in calculation of $\mathbb{N}$ (\ref{eq:N}),
as being $\left(kL\right)^{-1}$ smaller than the similar terms associated
with the short-range part of the full correlator $K$.\end{thebibliography}
\end{document}